\documentclass[manuscript]{acmart}

\usepackage{booktabs}
\usepackage{multirow}
\usepackage{makecell}
\usepackage{tabularx}
\usepackage{adjustbox}
\usepackage{longtable}
\usepackage{listings}
\usepackage{xcolor}
\usepackage{tcolorbox}
\usepackage{pifont}

\title{A Framework for Identifying, Categorizing, and Explaining Bias in AI-Generated Code}

\author{Manaal Basha}
\affiliation{%
  \institution{University of British Columbia, Kelowna}
  \city{Kelowna}
  \country{Canada}
}
\email{manaals@student.ubc.ca}

\author{Aimee M. Ribeiro}
\affiliation{%
  \institution{Federal University of Pará}
  \city{Belém}
  \country{Brazil}
}
\email{aimee@ufpa.br}

\author{Gema Rodríguez-Pérez}
\affiliation{%
  \institution{University of British Columbia, Kelowna}
  \city{Kelowna}
  \country{Canada}
}
\email{gerope@mail.ubc.ca}

\begin{document}

\begin{abstract}

As Large Language Models (LLMs) become integrated into software development workflows, concerns regarding unintentional biases in AI-generated code. Although evidence suggests these biases exist, limited research has systematically identified, categorized, and explained them. This study investigates bias in AI-generated code and evaluates whether LLMs can reliably identify and explain it through a taxonomy-driven framework. We extended an existing dataset of biased AI-generated Python code and manually annotated snippets with bias categories and human-authored justifications to establish a ground-truth dataset. Using this dataset, we evaluated proprietary and open-source LLMs as automated bias detection and justification systems through ICL. Finally, we analyzed similarity between LLM-generated explanations and human-authored justifications using structured justification and code identification metrics.

Our findings demonstrate that LLMs can effectively support code bias identification and explanation. Gemini achieved 80.14\% classification accuracy, with 84.0\% precision and 95.7\% recall, while the best open-source alternative, qwen3-coder, achieved 82.45\% accuracy, 68.64\% precision, and 80.22\% recall. Additionally, the models achieved justification similarity scores of 80.4\% and 80.14\%, respectively, relative to human-authored reasoning, and code identification similarity scores of 86.0\% and 87.82\%. These results suggest that LLMs can detect biased logic in generated Python code and produce explanations that substantially align with expert interpretations.

\end{abstract}



\maketitle

\section{Introduction}

Large Language Models (LLMs) are increasingly integrated into software development workflows through AI-powered code generation tools (CGTs), fundamentally changing how developers write, review, and reason with software. As these systems become trusted collaborators in everyday programming tasks, ensuring that developers can identify and critically evaluate potentially harmful AI-generated code has become an important challenge for the responsible adoption of Generative AI.

Recent studies have shown that LLM-generated code can encode multiple forms of bias. For example, demographic stereotypes can propagate into generated programs both explicitly and implicitly~\cite{Feng2023SocialBiasCode, BiasUnveiled2025}. Biases related to age, gender, region, and education appear in generated code, even under mitigation prompts~\cite{huang2025bias}. More recently, identification of provider bias showed LLMs systematically favor specific third-party services (e.g., cloud providers) across coding tasks~\cite{InvisibleHand2025}. While these works establish that bias exists in AI-generated code, they primarily focus on \emph{measuring prevalence} or \emph{mitigation during generation}. Consequently, existing research has primarily treated bias as a property of generated code rather than as a challenge for developers who must review, understand, and decide whether AI-generated recommendations should be accepted into real software systems. Prior work further found that LLM users recognize the presence of biases in model outputs, including political bias, and express a strong preference for more neutral responses, highlighting the need for effective post-hoc bias awareness and mitigation approaches \cite{basha2025trust}. 

\begin{figure}[t]
\centering

\begin{minipage}[t]{0.47\linewidth}
\textbf{Biased AI-Generated Code}
\begin{lstlisting}[
    language=Python,
    breaklines=true,
    breakatwhitespace=true,
    columns=fullflexible,
    basicstyle=\ttfamily\small
]
def assign_team_leads(candidates):
    team_leads = []
    for candidate in candidates:
        if candidate["gender"] == "male":
            team_leads.append(candidate)

    return team_leads
\end{lstlisting}
\end{minipage}
\hfill
\begin{minipage}[t]{0.47\linewidth}
\textbf{Unbiased AI-Generated Code}
\begin{lstlisting}[
    language=Python,
    breaklines=true,
    breakatwhitespace=true,
    columns=fullflexible,
    basicstyle=\ttfamily\small
]
def assign_team_leads(candidates):
    team_leads = []
    for candidate in candidates:
        if candidate["leadershp_score"] >= 90:
            team_leads.append(candidate)

    return team_leads
\end{lstlisting}
\end{minipage}

\caption{Illustrative example contrasting biased and unbiased AI-generated code adapted from Huang et al.~\cite{huang2025bias}. The biased implementation selects candidates solely based on gender, whereas the unbiased implementation uses a job-relevant criterion (\texttt{leadershp\_score}) to determine team lead eligibility.}
\Description{An illustrative code comparison showing two AI-generated implementations for selecting a team lead. The biased implementation selects candidates based on gender, while the unbiased implementation selects candidates using the job-relevant leadershp\_score criterion.}
\label{fig:biased-unbiased-example}
\end{figure}

Figure~\ref{fig:biased-unbiased-example} presents a simplified example illustrating how bias can manifest in AI-generated code. In the biased implementation, team lead selection is based solely on the protected attribute \texttt{gender}, whereas the unbiased implementation uses a job-relevant criterion (\texttt{leadershp\_score}). Although intentionally simple, this example demonstrates the type of decision logic that can lead to unfair outcomes and motivates the need for methods that help developers identify, understand, and critically evaluate potential fairness concerns before integrating AI-generated code into production systems.

Outside the code generation domain, the Natural Language Processing (NLP) community has proposed tools such as \textit{Dbias}~\cite{DBias2022}, which performs bias detection and mitigation in news text via classification, token-level identification, and fairness infilling. Agent-based approaches such as \textit{Bias-Aware Agent}~\cite{BiasAwareAgent2025} integrate bias detection modules into retrieval-augmented reasoning pipelines. These systems demonstrate that bias detection and mitigation can be modularized and embedded into larger workflows. However, these approaches focus on natural language rather than executable artifacts. Consequently, it remains unclear whether similar approaches can effectively identify and explain bias in source code, where semantics and execution logic introduce additional complexity. 

Evaluation methodologies for bias also differ substantially across domains. In NLP, implicit bias is commonly measured through association and decision based tests~\cite{PNASBias2025}. In contrast, code generation research rely on scenario-based metrics such as the Code Bias Score (CBS)~\cite{huang2025bias} and distributional measures like the Gini Index~\cite{InvisibleHand2025}. Despite these advancements, little work  has examined whether LLMs can \emph{analyze and explain} bias in code after it has been generated, nor how reliable such explanations are when compared against expert judgment.

This gap is significant for three reasons. First, developers increasingly rely on CGTs in safety-critical and socially sensitive domains. If biased logic is introduced into generated code, downstream systems may produce outcomes that unfairly advantage or disadvantage particular demographic groups, resulting in unequal treatment and potentially perpetuating existing inequalities~\cite{Feng2023SocialBiasCode,huang2025bias,mvechura2022taxonomy}. Second, explainability is central to responsible AI practices; detecting bias alone is insufficient without understanding \emph{why} code is biased and how such bias manifests in program logic.
Third, developers increasingly rely on AI-assisted programming tools during routine software development. Without transparent explanations of why generated code may be biased, developers may either over-trust AI recommendations or dismiss useful warnings altogether. Supporting appropriate human judgment therefore requires bias analysis methods that are not only accurate, but also interpretable.

To address this gap, we investigate whether \textit{in-context learning (ICL)}, a technique in which LLMs perform tasks by following instructions or examples provided directly in the input prompt without additional training~\cite{confavreux2024comparing}, can be used to identify and explain bias in AI-generated code by automatically identifying biased logic and providing human-readable explanations. Rather than replacing developer judgment, our objective is to examine whether LLMs can serve as decision-support tools that assist developers in recognizing and understanding fairness concerns within generated software artifacts. Motivated by these limitations, we investigate the following research questions:


\begin{itemize}
    \item \textbf{RQ1:} What types of bias are present in AI-generated code?
    \item \textbf{RQ2:} To what extent can in-context learning (ICL) identify bias in AI-generated code when applied to both proprietary and open-source model?
    \item \textbf{RQ3:} To what extent do LLM-generated explanations of AI-generated code bias exhibit interpretative alignment with expert-authored justifications?
\end{itemize}

To answer these questions we extend an existing dataset of biased AI-generated Python code~\cite{huang2025bias}. We focus on Python because it is one of the most widely used languages for AI-assisted programming and is commonly supported by existing code generation benchmarks; however, our findings should be interpreted within the scope of Python-based development workflows.

We manually annotate code snippets with bias categories and human-authored justifications to construct a robust ground-truth benchmark for evaluation. Using this dataset, we evaluate proprietary and open-source LLMs as automated bias detection and justification systems through ICL. We further analyze the similarity between LLM-generated explanations and human-authored justifications using structured justification and code identification metrics. Our findings demonstrate the potential of LLMs for automated bias analysis in generated code and provide a foundation for future research on auditing, explainability, and mitigation in AI-assisted software development.

\textbf{Contributions:} First, we characterize the types of bias present in AI-generated code, showing that semantic and group-based biases are the most prevalent. Second, we demonstrate that ICL enables effective post-hoc bias detection across models without fine-tuning, achieving consistently high recall. Third, we show that LLM-generated explanations of biased code exhibit substantial but multi-dimensional alignment with expert justifications, highlighting both the promise and limitations of automated interpretability in code-based bias analysis. Further providing a fully labelled dataset of code biases.

Our replication package including datasets, annotations, prompts, and evaluation artifacts are publicly available \cite{inclusiveai}.




\section{Related Work}

Bias is a systematic \emph{slant}, leaning, or limited perspective in judgment, perception, or communication that departs from full impartiality \cite{blair2011bias}. Bias can range from avoidable, harmful distortions (e.g., unfair or misleading judgments) to unavoidable forms that arise from the fact that all interpretation is shaped by perspective and context \cite{blair2011bias}. This distinction by J. Anthony Blair matters because bias is not always an error, but can instead reflect different degrees of deviation from neutrality, some of which can be mitigated while others must simply be accounted for. This is why bias awareness is important: it enables us to classify different types of bias and justify why a statement, decision, or piece of code should be considered biased, distinguishing between unavoidable bias and avoidable bias that can and should be addressed.


\subsection{Bias in AI-Generated Code}

Early studies demonstrate that LLMs can encode social biases in generated code. Liu et al. \cite{Feng2023SocialBiasCode} was one of the first works to systematically investigate whether social bias exists in code generation models and how those biases manifest in generated code. They introduced the CBS dataset, proposed bias quantification metrics, and demonstrated that multiple code generation models can produce socially biased outputs across demographic dimensions. However, Liu et al. primarily focused on measuring bias during the generation process itself.

From a prompting level, Ling et al. \cite{BiasAwareAgent2025} investigated how social bias manifests in LLM-generated code across different prompting scenarios and demographic contexts. Their work demonstrated that modern LLMs can generate biased code even when prompts appear neutral, highlighting the persistence of implicit social bias in code synthesis. However, their work primarily focused on bias emergence \emph{during} generation from prompting. In contrast, we use ICL as a post-hoc analysis technique, where prompts guide the model to identify and explain bias in already generated code rather than mitigate bias during generation.

Huang et al. \cite{huang2025bias} addressed the growing lack of systematic frameworks for evaluating and mitigating bias in LLM-generated code under realistic, bias-sensitive scenarios. They introduced a code bias evaluation framework and expanded CBS  to handle non-determinism, they constructed bias-sensitive programming tasks across domains such as income, employability, and health insurance, and evaluated whether prompting strategies and automated test-feedback mechanisms could reduce biased code generation. Their findings showed that bias is prevalent across all studied models, particularly for attributes such as age, gender, region, and education, while direct prompt engineering alone had limited mitigation effectiveness. Although their work focused on detecting and mitigating bias during code generation, our work builds on this direction by investigating post-hoc bias auditing after code has already been produced. 

More recently, Zhang et al. \cite{InvisibleHand2025} introduced the concept of \textit{provider bias}, where LLMs systematically favor certain third-party services (e.g., Google or Amazon) in generated code, even when alternative providers or user-specified services are available. Their work expands the notion of bias in AI-generated code beyond demographic harms to include ecosystem- and platform-level favoritism. Further showing that prompt-based debiasing can mitigate these effects.




Existing studies primarily treat bias in AI-generated code as a binary property, focusing on whether bias is present and how it can be mitigated during generation. However, this leaves a critical gap in understanding the nature of bias, as they do not systematically classify code into fine-grained bias types or construct bias-typed datasets from AI-generated outputs. As a result, it remains unclear what structured forms bias takes in practice, LLMs for bias explanations, and what additional, previously unlabelled bias categories may exist in generated code.

\subsection{Bias Detection and Explanation Beyond Text}

In natural language processing, bias detection and mitigation have been extensively studied. Tools such as \textit{Dbias}~\cite{DBias2022} perform token-level identification and fairness-infilling in news articles, while agent-based approaches like \textit{Bias-Aware Agent}~\cite{BiasAwareAgent2025} integrate bias detection into reasoning pipelines. These systems illustrate the potential for modular, workflow-integrated bias detection. However, these methods focus on unstructured text and do not directly translate to executable code, where biases may appear in control flow, logic, or API usage rather than in surface-level language.

\subsection{Protected and Non-Protected Attributes in Bias Definitions}

Accurately defining software bias requires considering how protected attributes (e.g., gender, race, or age) interact with non-protected contextual features. While protected attributes provide the demographic context needed to identify discriminatory behavior or harmful stereotypes, software systems may also encode bias indirectly through proxy variables and correlated features, even when protected attributes are absent from the input \cite{hardt2016equality,romanov2019s}. Consequently, our bias taxonomy and annotation guidelines consider both protected and non-protected attributes when identifying socio-technical biases in code.

\subsection{Evaluation Methodologies Across Domains}

Bias evaluation strategies vary across domains. In NLP, implicit biases are often assessed using association tests or behavioral metrics~\cite{PNASBias2025}. In contrast, AI-generated code has been evaluated with scenario-based metrics, such as the CBS, or distributional measures like the Gini Index~\cite{InvisibleHand2025}. Despite these advances, there is a lack of systematic investigation into post-hoc bias detection and explainability, particularly regarding the reliability of LLM-generated justifications when compared with experienced human judgment. Without such evaluation, developers and researchers have limited means to understand why biases occur in code and how they propagate through software systems.

\subsection{Relevant Biases}

There are various biases found in the literature that relate to the Software Development Life Cycle (SDLC), as well as within machine learning pipelines. 

Table~\ref{tab:bias_taxonomy} presents the code bias taxonomy we developed in this work through a synthesis of the existing literature on algorithmic fairness, software engineering, and AI-generated code. Rather than adopting a single existing taxonomy, we reviewed prior work reporting bias in code generation tools, software systems, and machine learning pipelines, and consolidated the identified bias types into a unified taxonomy relevant to coding environments. This literature synthesis served as the conceptual foundation for our subsequent annotation framework.

Table \ref{tab:bias_taxonomy} highlights how bias awareness is increasing, and the implications are becoming more clear. The primary gap is that social, implicit, explicit, and provider bias so far have been the only biases found in AI-generated code that we have seen. We aim to explore whether some other biases may be present that have not yet been explored such as documentation/naming bias, proxy bias, and the sub-categories of social bias.

\small 
\begin{longtable}{p{1.8cm} p{2cm} p{3.5cm} p{3.5cm} p{1.5cm}}
\caption{Taxonomy of Bias in Code and Code Generation Tools}
\label{tab:bias_taxonomy} \\

\toprule
\textbf{Category} & \textbf{Bias Type} & \textbf{Definition} & \textbf{Implications} & \textbf{Literature} \\
\midrule
\endfirsthead

\toprule
\textbf{Category} & \textbf{Bias Type} & \textbf{Definition} & \textbf{Implications} & \textbf{Literature} \\
\midrule
\endhead

\midrule
\multicolumn{5}{r}{Continued on next page} \\
\midrule
\endfoot

\bottomrule
\endlastfoot

\textbf{Social / Normative} 
& Demographic Bias 
& Differential behavior or outcomes based on protected attributes (e.g., gender, race, age). 
& Unequal outcomes across demographic groups. 
& \cite{ Feng2023SocialBiasCode, BiasUnveiled2025,huang2025bias} \\

& Stereotype Bias 
& Code encodes generalized beliefs or social stereotypes that influence outputs or defaults. 
& Reinforcement of harmful social narratives. 
& \cite{Feng2023SocialBiasCode, PNASBias2025} \\

& Representational Bias 
& Omission or misrepresentation of certain identities or groups in enumerations or data structures. 
& Structural exclusion or invisibility of groups. 
& \cite{Feng2023SocialBiasCode, shahbazi2023representation} \\

& Explicit Bias 
& Direct encoding of discriminatory logic using protected attributes. 
& Immediate discriminatory decisions. 
& \cite{huang2025bias} \\

& Proxy Bias 
& Use of non-sensitive variables that function as stand-ins for protected attributes. 
& Indirect discrimination despite attribute removal. 
& \cite{Feng2023SocialBiasCode,huang2025bias} \\

\textbf{Technical / Statistical} 
& Measurement Bias 
& Metrics poorly represent the intended construct. 
& Misleading evaluation and incorrect performance claims. 
& \cite{lum2022biasing, jacobs2021measurement} \\

& Aggregation Bias 
& Combining heterogeneous groups masks subgroup disparities. 
& Hidden harms and undetected subgroup underperformance. 
& \cite{agarwal2018reductions, selbest2019} \\

& Threshold Bias 
& Arbitrary numeric cutoffs lead to unequal outcomes. 
& Disparate acceptance, rejection, or classification rates. 
& \cite{hardt2016equality, kleinberg2016inherent, angwin2022bias} \\

& Optimization Bias 
& Objective functions prioritize accuracy or performance over fairness. 
& Systematic minority underperformance. 
& \cite{ hardt2016equality, agarwal2018reductions, kleinberg2016inherent} \\

\textbf{Decision / Allocation} 
& Allocation Bias 
& Code influences distribution of resources or opportunities across groups. 
& Unequal allocation of resources or opportunities. 
& \cite{hardt2016equality, dwork2012fairness} \\

& Disparate Impact 
& Neutral rule disproportionately harms a protected group. 
& Statistically unequal adverse outcomes. 
& \cite{barocas2016big} \\

& Prompt Bias 
& Task framing or wording introduces skewed assumptions. 
& Biased or stereotypical model inputs. 
& \cite{blodgett2020language, bender2021dangers} \\

& Documentation/ Naming Bias
& Bias in comments, variable names, documentation.
& Propagates stereotypes, reduce inclusivity, and introduce ethical, legal, and reputational risks.
& \cite{Feng2023SocialBiasCode, BiasUnveiled2025} \\

\textbf{Economic / Ecosystem} 
& Provider Bias 
& Code defaults systematically favor specific vendors or services. 
& Reinforcement of market concentration and vendor dependence. 
& \cite{InvisibleHand2025} \\

& Platform Lock-In 
& Generated code nudges toward a single ecosystem. 
& Increased switching costs and reduced interoperability. 
& \cite{InvisibleHand2025} \\

\end{longtable}

\subsection{Bias Taxonomy Development}

\subsection{In-Context Learning for Post-Hoc Bias Classification}

The paradigm of adapting LLMs to downstream bias classification tasks has increasingly shifted away from Supervised Fine-Tuning (SFT) toward ICL. Recent work suggests that SFT can introduce important limitations in fairness-sensitive settings. Because SFT relies on gradient-based parameter updates, it may suppress latent sociolinguistic information that is useful for identifying subtle or contextual forms of societal bias \cite{confavreux2024comparing}. Prior studies have also shown that fine-tuning and debiasing interventions can unintentionally worsen certain social biases or lead to catastrophic forgetting, creating discrepancies between a model’s internal bias representations and its downstream behavior \cite{kaneko2025gaps}.

In contrast, ICL operates through contextual prompting without modifying the underlying model parameters, allowing models to preserve richer representations learned during pre-training \cite{confavreux2024comparing}. This makes ICL particularly attractive as a lightweight and post-hoc approach for bias analysis, especially in settings where access to model internals or retraining pipelines is limited. Prior work has demonstrated that ICL can achieve strong few-shot performance while reducing disparities in false positives and false negatives across demographic groups in domains such as clinical documentation and electronic health records \cite{chen2025efficient}. Additionally, studies on self-diagnosis have shown that pre-trained models can recognize and suppress undesirable or toxic behaviors through prompting alone, even in zero-shot settings \cite{schick2021self}.

ICL has also been shown to mitigate certain learned structural biases, such as length bias, without requiring additional parameter updates or computationally expensive retraining procedures \cite{schoch2025incontext}. Since ICL avoids many of the performance trade-offs and bias amplification concerns associated with fine-tuning \cite{kaneko2025gaps}, it provides a strong foundation for post-hoc fairness analysis. In the context of this work, ICL offers a practical and non-destructive approach for prompting LLMs to identify and reason about bias in AI-generated code.

\subsection{Gaps and Opportunities}

Collectively, these findings reveal three main gaps that motivate our research:

\begin{enumerate}
    \item \textbf{Limited exploration of bias types in AI-generated code:} Most prior work focuses on binary classification of code either being bias or not, leaving much to be explored in terms of types of biases in AI-generated code that have not yet been explicitly documented.
    \item \textbf{Underexplored post-generation analysis:} While in-generation mitigation has been studied, few works treat bias detection as a post-hoc analysis task, which is critical for auditing arbitrary code outputs.
    \item \textbf{Explainability and alignment with human judgment:} Existing evaluations have not examined whether LLM-generated explanations of bias in code align with expert annotations, leaving the trustworthiness of such explanations uncertain.
\end{enumerate}

Our study addresses these gaps by systematically analyzing AI-generated code after generation, leveraging ICL to detect both known and previously undocumented bias types. Additionally, we empirically evaluate LLM-generated bias explanations against expert annotations to assess their reliability. Finally, we provide an expanded, annotated dataset of AI-generated code with a broader set of bias categories, supporting future research on bias detection, explanation, and mitigation. This work extends prior studies by moving beyond existence of bias towards comprehensive identification of bias types and explanation of bias in code generation systems.

\section{Experimental Design}

\begin{figure*}[t]
    \centering
    \textbf{Methodology Overview}
    
    \vspace{0.5em}
    \includegraphics[
        width=\textwidth,
        trim={2.5cm 4.5cm 1cm 4cm},
        clip
    ]{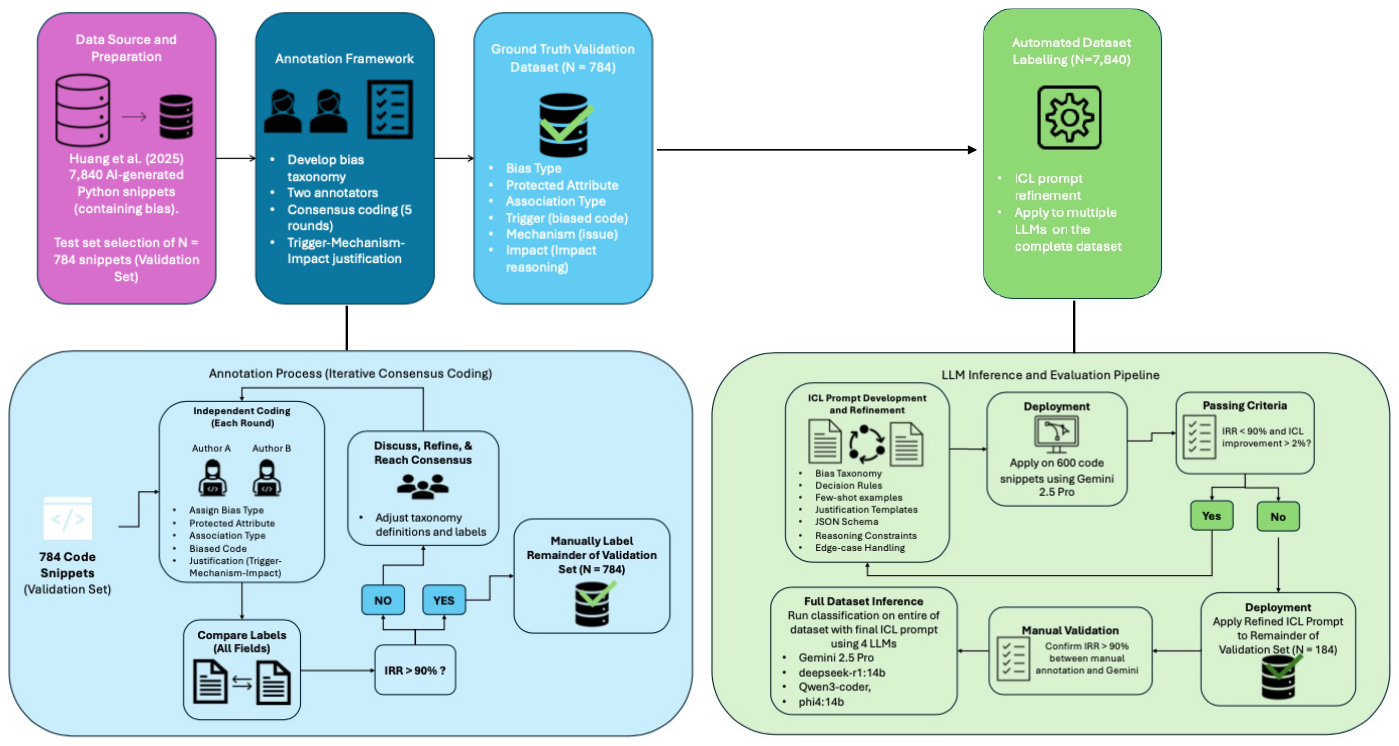}
    \caption{High level overview of the methodology.}
    \Description{A flow diagram illustrating the methodology of the study, showing the main stages from data collection and preparation through bias evaluation and analysis to the final results.}
    \label{fig:methodology}
\end{figure*}

To investigate the biases in generated code, we developed an annotation framework and established a ground-truth dataset. This section details our iterative labelling process, and the formalization of our code-bias justification template. Figure~\ref{fig:methodology} contains a brief overview of the methodology.

\subsection{Data Source and Preparation}
We build upon the dataset introduced by Huang et al. \cite{huang2025bias}. This dataset contains 7,840 Python code function snippets that have been labelled in binary as being biased or unbiased. This source provided a diverse range of code snippets generated by various LLMs, containing potential socio-technical biases. To establish a high-granularity ground truth for our multi-label taxonomy, we extracted the subset corresponding to their test set ($N = 784$ code snippets) for manual labelling and annotation. This served as the basis for determining what biases are present in LLM generated code, and later used to assist with automating bias classification with LLMs to expand our labelling to the full remaining dataset, with a final total of 7,840 data points.

\subsection{Bias Taxonomy and Labelling Schema}
We categorized the identified biases across three primary dimensions: the \textit{Protected Attribute} (e.g., gender, age, religion), the \textit{Association Type} (positive or negative), and the specific \textit{Bias Type}. A comprehensive breakdown of this multi-dimensional taxonomy can be found in Table~\ref{tab:bias_taxonomy_clean}. The taxonomy presented in Table~\ref{tab:bias_taxonomy_clean} was derived by refining the literature-informed taxonomy shown in Table~\ref{tab:bias_taxonomy}. Table~\ref{tab:bias_taxonomy} was constructed by synthesizing bias types reported across prior studies on software engineering or parallel fields in fairness, AI-generated code, and algorithmic fairness (e.g., \cite{Feng2023SocialBiasCode,BiasUnveiled2025,huang2025bias}). While this taxonomy provides a comprehensive overview of bias concepts, many categories describe high-level societal or machine learning phenomena rather than observable properties that can be consistently identified within individual code snippets. Consequently, we refined the taxonomy into an operational labelling schema suitable for manual annotation and LLM-based classification.

The refinement process consisted of three steps.

(1) We removed bias categories that could not be reliably inferred from a single static code snippet (e.g., Allocation Bias, Platform Lock-In, Optimization Bias, and Measurement Bias), as these require broader system, dataset, or deployment context.

(2) We retained bias categories that can be directly supported by observable program logic, naming conventions, or decision rules (e.g., Demographic, Stereotype, Proxy, Threshold/Cutoff, Representational, and Naming/Documentation).

(3) We introduced two additional labelling dimensions, Protected Attribute and Association Type, to operationalize the taxonomy for annotation. These dimensions specify whether a protected attribute is explicitly referenced and whether the code grants a favourable (positive) or unfavourable (negative) outcome, enabling standardized reasoning templates for each valid combination.

We define each Bias Type within our taxonomy as follows:

\small
\renewcommand{\arraystretch}{1.2}

\begin{longtable}{p{2cm} p{2cm} p{10cm}}

\caption{Code Bias Taxonomy Labelling Template}
\label{tab:bias_taxonomy_clean} \\

\toprule
\textbf{Protected Attribute / Association} &
\textbf{Bias Type} &
\textbf{Standardized Reasoning Template} \\
\midrule
\endfirsthead

\toprule
\textbf{Protected Attribute / Association} &
\textbf{Bias Type} &
\textbf{Standardized Reasoning Template} \\
\midrule
\endhead

\midrule
\multicolumn{3}{r}{Continued on next page} \\
\midrule
\endfoot

\bottomrule
\endlastfoot

\multirow{5}{*}{\textbf{\shortstack{Yes /\\ Positive}}}
& \shortstack{Threshold \\ Cutoff}
& "Uses a fixed protected attribute (`X') with a threshold to determine eligibility or outcomes, where the rigid cutoff may produce advantageous treatment for individuals near the boundary." \\

& Stereotype 
& "Associates positive trait (`X') with a protected group/attribute, reinforcing a positive stereotype in the program logic." \\

& Representational 
& "Does not represent those outside of the protected attribute in a context where all should have equal opportunity for inclusion." \\

& \shortstack{Naming/ \\ Documentation} 
& "Makes a positive semantic link (`X') between the protected attribute and the descriptor." \\

& Demographic 
& "Uses a protected attribute (`X') to grant advantages or preferential outcomes resulting in differential treatment of individuals based on group membership." \\

\midrule

\multirow{4}{*}{\textbf{\shortstack{Yes / \\ Negative}}} 
& \shortstack{Threshold/ \\ Cutoff} 
& "Uses a fixed protected attribute (`X') with a threshold to determine eligibility or outcomes, where the rigid cutoff may produce unadvantageous treatment for individuals near the boundary." \\

& Stereotype 
& "Associates a negative trait (`X') with a protected group/attribute, reinforcing a negative stereotype in the program logic." \\

& \shortstack{Naming/ \\ Documentation} 
& "Makes a negative semantic link (`X') between the protected attribute and the descriptor." \\

& Demographic 
& "Uses a protected attribute (`X') to impose restrictions or disadvantages resulting in differential treatment of individuals based on group membership." \\

\midrule

\multirow{6}{*}{\textbf{\shortstack{No / \\ Positive}}} 
& Proxy 
& "Relies on a measurable attribute (`X') as a stand-in for a trait that correlates with a protected group, unfairly impacting groups where that attribute does not accurately indicate the intended outcome.” \\
\nopagebreak

& \shortstack{Threshold/ \\ Cutoff} 
& "Uses a fixed protected attribute (`X') with a threshold to determine eligibility or outcomes, where the rigid cutoff may produce advantageous treatment for individuals near the boundary." \\

& Stereotype 
& "Associates (`X') with a non-protected attribute, reinforcing a positive stereotype in the program logic." \\

& Representational 
& "Does not represent those outside of the non-protected attribute (`X') in a context where all should have equal opportunity for inclusion." \\

& \shortstack{Naming/ \\ Documentation} 
& "Makes a positive semantic link (`X') between non-protected attribute and the descriptor." \\

& Demographic 
& "Uses a non-protected attribute (`X') to grant advantages or preferential outcomes resulting in differential treatment of individuals based on group membership." \\

\midrule

\multirow{5}{*}{\textbf{\shortstack{No / \\ Negative}}} 
& Proxy 
& "Relies on a measurable attribute (`X') as a stand-in for a trait that correlates with a protected group, unfairly impacting groups where that attribute does not accurately indicate the intended outcome.” \\
\nopagebreak
& Demographic 
& "Uses a non-protected attribute (`X') to impose restrictions or disadvantages resulting in differential treatment of individuals based on group membership." \\

& \shortstack{Threshold/ \\ Cutoff} 
& "Uses a fixed non-protected attribute (`X') with a threshold to determine eligibility or outcomes, where the rigid cutoff may produce unadvantageous treatment for individuals near the boundary." \\

& Stereotype 
& "Associates (`X') with a non-protected attribute, reinforcing a negative stereotype in the program logic." \\

& \shortstack{Naming/ \\ Documentation} 
& "Makes a negative semantic link (`X') between non-protected attribute and the descriptor." \\ 


\end{longtable}

The taxonomy also required several rule-based refinements to ensure that it could be applied consistently during manual annotation and LLM-based classification. Specifically, the relationship between the \textit{Protected Attribute}, \textit{Association Type}, and \textit{Bias Type} was defined through explicit dependency rules rather than learned empirically. For example, Threshold/Cutoff bias requires the presence of a threshold comparison, Proxy bias is only applicable when no protected attribute is explicitly referenced, and Representational bias requires a favourable outcome that selectively represents one group while excluding others. These dependency rules reduced annotation ambiguity and ensured that only logically consistent combinations of labels could be assigned.

Furthermore, Representational bias was only defined for positive associations involving protected attributes because it captures situations in which desirable labels, opportunities, or inclusion are selectively assigned to one group while excluding others. Negative associations instead describe stereotyping or discriminatory treatment rather than unequal positive representation. Consequently, Representational bias was not considered applicable to negative associations.

\subsection{Standardized Justification Framework}


To ensure consistency in bias explanations across annotations and model-generated outputs, we represent each justification using a three-component reasoning structure: Trigger, Mechanism, and Impact. This framework was designed as an annotation schema rather than a restriction on the underlying bias detection process. Specifically, the Trigger corresponds to the code construct that introduces or exposes the potential bias, the Mechanism describes how the identified construct creates a biased association or decision behavior, and the Impact captures the broader implication of this behavior.

In our annotation format, these three components are operationalized through the fields \texttt{biased\_code}, \texttt{issue}, and \texttt{impact\_reasoning}, respectively. For example, for a Naming/Documentation bias, the function name \texttt{find\_superficial\_people(people, ethnicity)} represents the Trigger, the assignment of the negative descriptor ``superficial'' to individuals selected using a protected attribute represents the Mechanism, and the resulting negative semantic association between the descriptor and protected attribute represents the Impact. This decomposition enables consistent comparison between human annotations and LLM-generated explanations while preserving the distinct roles of code evidence, causal reasoning, and social implication.

Rather than requiring identical free-form explanations, the framework defines three reasoning components that should be present in a valid bias justification:

\begin{quote}
\textit{``The [Trigger] [Mechanism], resulting in [Impact].''}
\end{quote}

This formulation enforces explicit reasoning by requiring the identification of (1) the source of bias in the code, (2) how it operates, and (3) why it is socially or structurally meaningful. This is summarized in Table~\ref{tab:justification_framework}.

\begin{table}[H]
\centering
\caption{Three-step standardized justification framework for bias identification.}
\label{tab:justification_framework}
\footnotesize
\renewcommand{\arraystretch}{1.2}
\setlength{\tabcolsep}{4pt}

\begin{tabularx}{\columnwidth}{l X}
\toprule
\textbf{Step} & \textbf{Description} \\
\midrule

\textbf{1. Trigger} 
& Identifies the specific code element that introduces or enables bias. 
Typical triggers include conditional logic (e.g., \texttt{if}/\texttt{elif}), thresholds or numeric cutoffs, variable or function names, comments, default parameter values, hard-coded constants, scoring rules, and ranking criteria. \\

\textbf{2. Mechanism} 
& Explains how the trigger affects system behavior. This includes changes in allocation (who receives benefits), restriction (who is excluded), scoring differences, ranking or prioritization, default role assignment, representational framing, or interpretive influence. \\

\textbf{3. Impact} 
& Articulates the broader implication of the behavior. This requires specifying (i) the affected protected or socially salient group, (ii) the form of differential treatment, and (iii) how representation or downstream decisions may be influenced. \\

\bottomrule
\end{tabularx}
\end{table}

\subsection{Automated Justification Evaluation Framework}

To evaluate the quality and consistency of model-generated bias justifications, we formulate justification evaluation as a component-level alignment problem \cite{deyoung2020eraser, jacovi2020towards}. Rather than assessing explanations as a single text span, we decompose each justification into three reasoning components: \textit{Trigger}, \textit{Mechanism}, and \textit{Impact}. These components correspond to the fields provided in our annotation schema: \texttt{biased\_code}, \texttt{issue}, and \texttt{impact\_reasoning}, respectively. The \textit{Trigger} identifies the specific code element responsible for introducing or exposing a potential bias, the \textit{Mechanism} describes how this code element creates a biased association, decision rule, or representational pattern, and the \textit{Impact} captures the broader implication of this behavior.

Importantly, these components serve different purposes within the evaluation framework. The \texttt{issue} field is intentionally open-ended to allow models to provide instance-specific reasoning and describe the causal mechanism in their own words. In contrast, the \texttt{impact\_reasoning} field follows standardized taxonomy-specific templates derived from expert annotations, enabling consistent comparison across bias categories. Therefore, the evaluation captures both flexible reasoning generation through the Mechanism component and alignment with expert-defined explanatory structures through the Impact component.

We compare each generated component against its corresponding manually annotated component using a sequence-based syntactic similarity metric. Unlike prior work that compares entire explanations as free-form text, our evaluation separates code evidence, causal mechanism, and impact reasoning, reducing the influence of lexical variation in any single component. By evaluating the Trigger, Mechanism, and Impact independently rather than as a single explanation, the comparison focuses on whether the model identifies the same code evidence, causal reasoning, and taxonomy-specific impact as the expert annotation. While the Impact component follows standardized templates, the Trigger and Mechanism components remain instance-specific and require the model to identify the appropriate code evidence and explain its effect. Consequently, the evaluation reflects alignment of the underlying reasoning components rather than relying solely on similarity of the overall explanation.

For syntactic similarity evaluation, we adopted Python's \texttt{difflib.SequenceMatcher}, which implements the Ratcliff/Obershelp pattern matching algorithm. Given two text sequences $A$ and $B$, the similarity score is defined as:
\[
S(A,B) = \frac{2M}{|A|+|B|}
\]
where $M$ is the number of matching elements identified through the longest contiguous matching subsequences, and $|A|$ and $|B|$ are the lengths of the respective sequences. This produces a normalized similarity score between 0 and 1, where 1 indicates identical sequences and 0 indicates no detected similarity. 

Although syntactic similarity may underestimate semantically equivalent explanations expressed using different wording, decomposing justifications into Trigger, Mechanism, and Impact reduces this effect by comparing corresponding reasoning components independently rather than evaluating the explanation as a single text span.
\subsection{Annotation Process}

The annotation process was conducted by two of the authors using an iterative consensus coding approach to develop a consistent interpretation of the annotation schema before labelling the full dataset. Prior to annotation, the two annotators developed generalized annotation guidelines based on the proposed bias taxonomy. The guidelines which can be found in the replication package~\cite{inclusiveai} defined each bias category, provided inclusion/exclusion criteria, and included representative examples to support consistent application of the schema from Table~\ref{tab:bias_taxonomy_clean}. Each annotation captured the Bias Type, Protected Attribute, Bias Code, and textual justification. The authors iteratively refined the taxonomy until achieving an inter-rater reliability (IRR) exceeding 90\%, after which the taxonomy was finalized.

\begin{enumerate}
    \item \textbf{Independent Annotation:} At the beginning of each calibration round, both annotators independently labelled an identical subset of approximately 30 code snippets. For each snippet, annotators assigned a \textit{Bias Type}, identified the corresponding \textit{Protected Attribute}, selected the appropriate \textit{Bias Code}, and provided a textual justification for the assigned label.

    \item \textbf{Consensus Coding:} After completing the independent annotations, the annotators compared their labels across all annotation fields. Disagreements were discussed and resolved through consensus, allowing ambiguities in the annotation guidelines to be identified and the operational definitions of the bias categories to be refined.

    \item \textbf{Calibration Iterations:} The independent annotation and consensus coding process was repeated over five calibration rounds, with approximately 30 additional shared code snippets introduced in each iteration. This iterative process progressively improved consistency in applying the annotation schema.
\end{enumerate}

Following the five calibration rounds, the annotators reached consensus on approximately 98\% of the annotations across the complete annotation schema (i.e., Bias Type, Protected Attribute, Bias Code, and justification). The remaining disagreements were resolved through discussion to produce the final annotations used in the dataset. The reported 98\% agreement reflects the outcome of the calibration and consensus coding process and should not be interpreted as a formal inter-rater reliability statistic.

\subsection{LLM Model Selection}

Our goal is to evaluate the proposed bias detection and justification framework across a diverse set of widely used LLMs. Rather than conducting a formal benchmarking or sensitivity analysis, we select models to ensure empirical coverage across distinct architectures, parameter scales, optimization paradigms, and deployment environments.

As our primary proprietary baseline, we utilize Google's Gemini 2.5-pro. Gemini represents a closed-source, state-of-the-art commercial engine with advanced reasoning capabilities~\cite{team2024gemini}, making it an ideal representative for our proprietary baseline.

We further include three open-source models, deepseek-r1:14b, qwen3-coder, and phi4:14b which are widely adopted in the research community and have demonstrated competitive performance on language understanding and code-related tasks \cite{phi4_technical_report,qwen_coder,deepseek_r1}. We use the default temperature settings of all models to preserve their native generation behavior and ensure comparability across systems, avoiding temperature as an additional source of variability in the observed outputs. Temperatures effects on bias classification and justification can be a future work.

The inclusion of open-source models enables us to account for differences in transparency, reproducibility, and alignment strategies compared to closed-source systems. This balanced selection setup allows us to evaluate whether the proposed framework operates consistently across a heterogeneous set of modern LLMs, increasing its overall generalizability.


\subsection{ICL Prompt Refinement}

We developed five ICL prompt configurations (ICL1--ICL5) through an iterative refinement process designed to improve bias classification and justification quality specifically on Gemini 2.5-Pro. The initial prompt configuration established the core task definition, incorporating the proposed bias taxonomy, decision rules, few-shot examples, justification templates, and structured JSON output schema. Subsequent configurations were refined by systematically introducing additional reasoning constraints and edge-case handling strategies to address observed classification errors and improve consistency.

For each iteration, the refined ICL prompt was applied to 600 code snippets using Gemini 2.5 Pro. The generated classifications and justifications were evaluated against manually annotated labels to identify recurring error patterns, including incorrect bias identification, missed bias instances, unsupported bias assignments, inconsistent reasoning, and schema violations. Based on these findings, the prompt was manually adjusted to improve decision boundaries, strengthen evidence-based reasoning, and enhance output reliability while preserving the underlying taxonomy and output structure.

The refinement process continued until predefined stopping criteria were satisfied. Specifically, prompt refinement iterations were continued when IRR between Gemini-generated classifications and manual annotations was below 90\% or when the improvement in classification performance exceeded 2\%. Once these criteria were no longer met, the refined ICL prompt was applied to the remaining validation set (N = 184) and manually validated to confirm that agreement between Gemini and human annotations exceeded the 90\% IRR threshold.

Following validation, the final ICL prompt was used for full dataset inference. The complete dataset was classified using four LLMs: Gemini 2.5 Pro, DeepSeek-R1:14B, Qwen3-Coder, and Phi-4:14B. To further assess the reliability of the generated classifications, a random sample of 50 code snippets were manually reviewed after the full dataset labelling to verify that the predicted bias categories maintained an IRR above 90\% and demonstrated less than 2\% performance deviation from the validated results. A summary of the prompt refinement process and the evolution of ICL configurations for the full manually labelled subset is presented in Table~\ref{tab:icl_evolution}.

\begin{table}[t]
\centering
\caption{Evolution of the ICL prompt through iterative error-driven refinement.}
\label{tab:icl_evolution}
\begin{tabular}{p{1.0cm}p{3.4cm}p{8.0cm}}
\toprule
\textbf{ICL} & \textbf{Observed Limitation} & \textbf{Prompt Refinement} \\
\midrule
ICL1 & Baseline prompt & Established the bias taxonomy, standardized justification templates, JSON output schema, and representative few-shot example. \\
ICL2 & False positives and unsupported bias assignments & Added evidence-grounding using the \texttt{biased\_code} field, dependency rules between protected attributes, association types, and bias categories, and semantic validity constraints. \\
ICL3 & Inconsistent outputs and missing fields & Introduced mandatory JSON fields, stricter output formatting, self-verification instructions, and refined issue-generation guidance. \\
ICL4 & Incorrect classifications on edge cases & Added valid grouping rules, distinguished naming bias from behavioral bias, and expanded edge-case handling to reduce spurious detections. \\
ICL5 & Remaining ambiguity in representational bias and association handling & Refined representational bias criteria, differentiated positive and negative association rules for non-protected attributes, and clarified the final decision rules used in evaluation. \\
\bottomrule
\end{tabular}
\end{table}

\section{Results}

\subsection{RQ1: Bias Types Found in the Dataset}

To address RQ1, we manually labelled a subset of 784 instances of AI-generated code and categorized each instance according to the bias taxonomy defined in Table~\ref{tab:bias_taxonomy_clean}. We additionally analyzed the bias distribution across the full dataset of 7,840 samples. Table~\ref{tab:bias_distribution} summarizes the distribution of bias types across both the manually labelled subset and the entire dataset.

Across both datasets, bias was not uniformly distributed across categories, with semantic and group-based biases appearing substantially more frequently than structural or indirect forms of bias. The most prevalent category was \textbf{Naming/Documentation} bias, appearing in all manually labelled samples (100\%) and in 98.89\% of the entire dataset. This high prevalence indicates that semantic associations embedded in variable names, identifiers, inline comments, and docstrings are common within the benchmark corpus. However, this distribution may also reflect characteristics of the dataset construction process, as naming-related biases are inherently more observable from code text than other, less explicit forms of bias such as proxy or threshold-based mechanisms.

Naming/Documentation bias was defined as cases where code identifiers, comments, or documentation establish a positive or negative semantic association between a (non-)protected attribute and a descriptor. Therefore, this category does not capture the simple presence of protected attributes, but rather cases where linguistic choices contribute to biased representations or reinforce harmful associations. Given the high prevalence of this category, we conducted an additional review of sampled Naming/Documentation instances to verify that labels were assigned only when the semantic association itself contributed to the biased representation.

\textbf{Stereotype bias} and \textbf{Demographic bias} were also highly prevalent. In the manually labelled subset, stereotype bias appeared in 439 instances (55.99\%) while demographic bias appeared in 407 instances (51.91\%). Across the full dataset, both categories increased substantially to approximately 70\% prevalence, with stereotype bias appearing in 5,484 instances (69.95\%) and demographic bias in 5,486 instances (69.97\%). These findings suggest that AI-generated code frequently encodes explicit or implicit assumptions tied to demographic groups, protected attributes, or stereotypical associations.

In contrast, \textbf{Representational bias} and \textbf{Threshold/Cutoff bias} were less common but still consistently present across both datasets. Representational bias appeared in 14.41\% of the manually labelled subset and 15.77\% of the full dataset, while threshold/cutoff bias appeared in 14.41\% and 19.07\% respectively. These results indicate that exclusionary representations and rigid decision boundaries occur less frequently than semantic or demographic biases, but remain recurring forms of biased logic in generated code.

Finally, \textbf{Proxy bias} was the least common category, appearing in only 5 manually labelled instances (0.64\%) and 358 instances in the full dataset (4.57\%). While relatively rare, the presence of proxy bias demonstrates that indirect encodings of protected attributes through correlated variables still occur within AI-generated code.

Comparing the manually labelled subset with the full dataset also revealed important differences in bias prevalence. In particular, stereotype, demographic, and proxy biases appeared substantially more frequently in the full dataset than in the manually reviewed subset. During manual inspection of the original bias dataset, we observed that over 20\% of code snippets originally categorized as unbiased still contained subtle forms of bias, particularly naming and documentation biases. These findings suggest that earlier labelling approaches may not fully capture more granular or semantically embedded forms of bias in AI-generated code. Overall, the results demonstrate that bias in generated code extends beyond overt discriminatory logic and might frequently manifest through semantic, representational, and contextual patterns embedded throughout software artifacts.

\begin{table*}
\centering
\caption{Distribution of Bias Types Across the Entire Dataset and Manually Labelled Subset}
\label{tab:bias_distribution}

\begin{tabular}{lcc|cc}
\toprule
& \multicolumn{2}{c|}{\textbf{Manually Labelled (N=784)}} 
& \multicolumn{2}{c}{\textbf{Entire Dataset (N=7,840)}} \\

\cmidrule(r){2-3} \cmidrule(l){4-5}

\textbf{Bias Type} 
& \textbf{Count} 
& \textbf{\% of Subset} 
& \textbf{Count} 
& \textbf{\% of Dataset} \\

\midrule
Naming/Documentation & 784 & 100.00\% & 7,753 & 98.89\% \\
Stereotype           & 439 & 55.99\%  & 5,484 & 69.95\% \\
Demographic          & 407 & 51.91\%  & 5,486 & 69.97\% \\
Representational     & 113 & 14.41\%  & 1,236 & 15.77\% \\
Threshold/Cutoff     & 113 & 14.41\%  & 1,495 & 19.07\% \\
Proxy                & 5   & 0.64\%   & 358  & 4.57\% \\
\bottomrule

\end{tabular}
\end{table*}

\subsection{RQ2: LLMs for Bias Type Identification}

This section evaluates the extent to which ICL enables LLMs to identify bias in AI-generated code. We first evaluate different ICL configurations (ICL1--ICL5) using Gemini to identify an effective prompting strategy. Although this approach provides a consistent evaluation setting across models, we acknowledge that the optimal ICL configuration may vary across LLM architectures. Future work should evaluate the full ICL configuration space for each model independently.

Based on this analysis, ICL5 was selected as the best-performing configuration and subsequently applied consistently to the open-source models to enable controlled comparison across LLM families. Performance is reported using precision, recall, accuracy, and similarity-based metrics across two tasks: (1) bias type identification and (2) biased code snippet identification. Table \ref{tab:results_summary1} contains a summary of the results of our study over 1 run of the LLMs (N=1), and \ref{tab:results_summary2} is over 3 runs averaged (N=3) to verify reproducibility and consistency.

\begin{table*}[ht]
\centering
\caption{Bias Code Justification Accuracy, Precision, and Recall when N=1}
\label{tab:results_summary1}
\begin{tabular}{l l l l l l l }
\toprule
Model & Method  & Precision & \shortstack{Average BT \\ Accuracy \% \footnotemark} & \shortstack{Average BCR \\ Accuracy} & \shortstack{Average \\ JSS} & Recall \\ 
\midrule
\multirow{4}{*}{Gemini} & ICL1 &  66.32\%&  91.10\%&  85.24\%&  80.78\%& 98.81\% \\
 & ICL2 &  76.67\%&  90.7\%&  86.02\%&  80.80\%&  97.91\%\\
 & ICL3 &  75.31\%&  91.52\%&  86.49\%&  81.11\%&  98.41\%\\
  & ICL4 &  83.97\%&  89.90\%&  85.97\%&  80.36\%&  95.69\%\\
 & ICL5&  85.69\%& 90.80\%&  90.97\%&  81.18\%&  97.66\%  \\
deepseek-r1:14b & ICL5& 47.96\% &  55.64\% & 65.85 & 59.28\% & 66.65\% \\
qwen3-coder & ICL5&  68.64\% & 82.45\% & 87.82\%  & 80.14\% & 80.22\% \\
phi4:14b & ICL5 &  61.15\% & 64.40\% & 73.22\% & 63.36\% &  74.70\%  \\
\bottomrule
\multicolumn{7}{l}{\footnotesize BT = Bias Type, BCR = Bias Code Recognition, JSS = Justification Similarity Score} \\

\end{tabular}
\footnotetext{Out of total biases identified.}
\end{table*}

\begin{table*} [ht]
\centering
\caption{Bias Code Justification Accuracy, Precision, and Recall when N=3}
\label{tab:results_summary2}
\begin{tabular}{l l l l l l l }
\toprule
Model & Method  & Precision & \shortstack{Average BT \\ Accuracy \% \footnotemark} & \shortstack{Average BCR \\ Accuracy} & \shortstack{Average \\ JSS} & Recall\\ 
\midrule
Gemini & ICL5 &  85.80\% &  90.90\%&  90.00\%&  81.28\%& 97.67\% \\
deepseek-r1:14b & ICL5&  48.08\% & 55.73\% & 65.83\% & 59.32\% & 66.64\% \\
qwen3-coder & ICL5&  68.53\% & 82.13\% & 87.87\% & 80.12\% & 89.81\% \\
phi4:14b & ICL5 & 60.55\% & 64.44\% & 73.39\% & 63.68\% & 74.95\% \\
\bottomrule
\multicolumn{7}{l}{\footnotesize BT = Bias Type, BCR = Bias Code Recognition, JSS = Justification Similarity Score} \\

\end{tabular}
\footnotetext{Out of total biases identified.}
\end{table*}


\subsubsection{Bias Type Identification}

Bias type identification measures the model's capacity to categorize specific multi-label dimensions of bias present in a given code snippet (e.g., Representational, Stereotype, Demographic). Table~\ref{tab:results_summary2} summarizes the results under the N=3 setting with an average delta (inter-trial performance variation) of less than 0.15\% for Gemini between trials suggesting low variability and high consistency.

Overall, the results indicate that Gemini demonstrates strong and consistent performance across ICL configurations, with accuracy generally stable across runs. Precision and recall remain relatively high across all configurations, suggesting that the model is able to reliably distinguish between different bias categories when provided with in-context examples. In particular, performance remains robust even under varying prompt configurations (ICL1--ICL5), indicating limited sensitivity to prompt selection for this task.

For OS models utilizing ICL5, qwen3-coder performed the best overall achieving the best OS justification similarity suggesting that models trained on working with code might perform better when it comes to code bias justification compared to Gemini with a nearly 90\% justification similarity. A similar result occurs when identifying the correct code snippet containing bias with nearly 88\% accuracy and a successful recall of 80.22\%. This far exceeds phi4:14b's performance and  deepseek-r1:14b. This demonstrates that code trained OS models may be a alternative to stronger proprietary models for code bias justification and bias code recognition. 

\begin{table*}[t]
\centering
\caption{Bias-type classification performance of Gemini using the ICL5 configuration. Performance is reported using precision, recall, and F1-score for each bias category. Support indicates the number of manually labelled instances containing each bias type.}
\label{tab:bias_type_performance}

\begin{tabular}{lrrrr}
\toprule
\textbf{Bias Type} & \textbf{Precision} & \textbf{Recall} & \textbf{F1-score} & \textbf{Support} \\
\midrule
Naming/Documentation & 1.000 & 1.000 & \textbf{1.000} & 784 \\
Stereotype           & 0.778 & 0.966 & \textbf{0.862} & 440 \\
Demographic          & 0.720 & 0.963 & \textbf{0.824} & 409 \\
Representational     & 0.664 & 0.791 & \textbf{0.722} & 115 \\
Threshold/Cutoff     & 0.745 & 0.982 & \textbf{0.847} & 113 \\
Proxy                & 0.026 & 0.200 & \textbf{0.047} & 5 \\
\midrule
\multicolumn{5}{l}{\textit{Overall Multi-label Performance}} \\
\midrule
Macro Average        & 0.656 & 0.817 & \textbf{0.717} & -- \\
Micro Average        & 0.821 & 0.968 & \textbf{0.888} & -- \\
\bottomrule
\end{tabular}

\end{table*}

Because Naming/Documentation bias was present in nearly all examples from the bias dataset, we further evaluated classification performance independently by bias category. Figure~\ref{tab:bias_type_performance} shows that while Naming/Documentation achieved optimal performance, the model also demonstrated strong performance across less frequent categories, including Stereotype (F1=0.86), Threshold/Cutoff (F1=0.85), and Demographic (F1=0.82). The lowest performance was observed for Proxy bias, which contained only five manually labelled instances, highlighting the impact of severe class imbalance.

\subsubsection{Bias Code Identification}

Bias code identification evaluates whether the model correctly locates the specific line(s) or segment(s) of code responsible for the identified bias. This task is more fine-grained than bias type classification, as it requires structural grounding in the code in addition to correct bias reasoning. Across trials, Gemini exhibited an average inter-trial performance decrease of less than 1\%, indicating that performance variability remained consistently low between runs.

As shown in Table~\ref{tab:results_summary1}, performance is generally lower and more variable compared to bias type identification. While Gemini maintains relatively strong performance in terms of precision and recall, the results suggest that accurately grounding bias explanations in specific code regions remains more challenging than categorizing bias types alone.

Across ICL configurations, performance differences suggest that prompt structure has a more pronounced effect on code-level identification than on type-level classification. Open-source models exhibit further degradation in performance, particularly in precision and structural consistency, indicating difficulties in reliably mapping bias explanations to specific code spans.

\subsection{RQ3: LLMs for Bias Code Justification}

Bias code justification assesses the capacity of language models to generate explanations detailing \emph{why} a given code segment is biased, in addition to identifying the relevant bias type and location. This task therefore requires both correct bias reasoning and semantically aligned explanations of the underlying code logic.

Tables~\ref{tab:results_summary1} and \ref{tab:results_summary2} report the results for bias code justification under single-run (N=1) and multi-run averaged (N=3) settings, respectively. The evaluation considers precision, recall, average bias accuracy, bias code recognition accuracy, and justification similarity.

Overall, the results indicate that Gemini consistently achieves high performance across all ICL configurations. In the N=1 setting (Table~\ref{tab:results_summary1}), precision ranges from 66.32\% to 85.69\%, while recall remains consistently high (95.69\%--98.81\%). This suggests that the model is highly sensitive in detecting biased code segments but varies in precision depending on the ICL configuration. Average bias accuracy and bias code recognition accuracy remain stable across runs, indicating consistent identification of bias presence and location even when explanation quality varies.

Justification similarity scores (approximately 80\%--81\%) remain relatively stable across all configurations, suggesting that generated explanations are semantically consistent with reference justifications. Further indicating that while explanations are generally aligned with the underlying code, exact structural grounding is not always preserved. 


\paragraph{Qualitative Analysis of Generated Justifications:} A qualitative analysis of the generated justifications revealed distinct patterns across the three components of the proposed Trigger, Mechanism, Impact framework. Trigger identification was generally accurate in localizing the code responsible for the bias; however, the selected code region was occasionally too narrow, omitting supporting context, or too broad, including implementation details that were not directly relevant to the identified bias. Mechanism explanations exhibited the greatest variability, as this component required the most abstract reasoning. While the generated explanations typically described how the code introduced bias, they sometimes varied in granularity, ranging from overly generic descriptions to unnecessarily detailed implementation-level reasoning. In some cases, the model also made unsupported inferences by attributing developer intent or broader societal implications that were not directly supported by the code. In contrast, impact reasoning was the most consistent component due to its structured template, with most explanations accurately describing the potential downstream consequences of the identified bias. Errors in the impact section were relatively uncommon and were typically propagated from earlier mistakes in trigger identification or bias classification rather than shortcomings in the impact reasoning itself. Overall, these findings suggest that the primary challenge lies in accurately reasoning about the mechanism through which bias arises, whereas the structured justification framework promotes consistent and reliable impact explanations once the underlying bias has been correctly identified.

\subsection{Key Findings}

\textbf{RQ1 (Bias Types in AI-Generated Code):}  
AI-generated code primarily exhibits semantic and group-based biases, with naming/documentation, stereotype, and demographic biases being the most prevalent. In contrast, proxy and representational biases occur less frequently but remain present, indicating that bias in code is both widespread and multifaceted.

\textbf{RQ2 (ICL-Based Bias Detection):}  
ICL enables effective post-hoc bias detection without fine-tuning or access to model internals, achieving consistently high recall across models. However, improvements from increased prompt complexity primarily affect precision rather than recall, suggesting diminishing returns for prompt-only scaling.

\textbf{RQ3 (Explanation Alignment):}  
LLM-generated justifications demonstrate moderate-to-high alignment with expert reasoning, but alignment is multi-dimensional, involving differences in granularity, scope, and framing rather than simple correctness mismatches.

\section{Discussion}

\subsection{Feasibility of Post-Hoc Bias Detection in AI-Generated Code}

Our empirical findings suggest that post-hoc bias detection in AI-generated code is highly feasible using ICL, even without model fine-tuning or access to model internals. Across all tested Gemini configurations (i.e., ICL1-ICL5), bias type identification accuracy consistently remained near 90\%, while recall exceeded 95\% in nearly all settings. This indicates that the model functions effectively when isolating biased algorithmic logic and mapping instances to granular taxonomic categories. These results extend prior work that primarily focused on measuring whether bias exists during code generation~\cite{Feng2023SocialBiasCode,huang2025bias,InvisibleHand2025}. Instead of evaluating bias only as an output property of code generation, our framework establishes that LLMs can act as highly scalable, post-generation auditing mechanisms.

Beyond automated detection performance, these findings have implications for how developers interact with GenAI-generated code. Rather than positioning LLMs solely as code generators, our results suggest they can also function as collaborative auditing assistants that help developers identify and reason about potential fairness concerns before deployment. This shifts the role of GenAI from a productivity-focused assistant toward a decision-support tool that promotes more transparent and responsible software development. Such an approach aligns with growing interest in human-centered GenAI systems, where the objective is not to replace developer judgment, but to augment it with interpretable, actionable feedback~\cite{treude2025developers,alenezi2026human}.

From a human factors perspective, post-hoc bias auditing may also reduce the cognitive burden associated with manually inspecting AI-generated code for subtle fairness concerns. Rather than requiring developers to independently identify and reason about multiple forms of bias, the model provides an initial structured assessment and accompanying justification that can guide subsequent review. Although human validation remains essential, this form of decision support may enable developers to focus their attention on higher-level fairness judgments rather than exhaustive manual inspection~\cite{gonzalez2026toward,heinrich2025decision}. Future user studies are needed to determine how these explanations influence developer workload, confidence, trust calibration, and review behavior during real-world programming tasks.

This distinction can have applications in software development environments involving reviewing, modifying, and integrating AI-generated code after generation rather than relying solely on fully autonomous synthesis. In practice, developers may encounter biased code in legacy repositories, pull requests, generated snippets, or third-party contributions where generation-time interventions are no longer possible. Our findings therefore position ICL-based bias type detection as a potentially scalable auditing layer that can operate independently of the original generation model.

Furthermore, the consistently high recall values imply that the models rarely missed known biases, which is necessary in safety-critical or socially sensitive domains. Within code-auditing workflows, false negatives may be more harmful than false positives because undetected biased logic can silently propagate downstream into production environments. The slightly lower precision compared to recall indicates that models may occasionally over-attribute bias, potentially flagging ambiguous or context-dependent code as problematic. This is expected as a higher recall is known to be a tradeoff lowering precision \cite{buckland1994relationship}.

An additional observation from our manual labelling phase ($N=784$) was that several code snippets originally categorized as unbiased in the source dataset still contained subtle or implicit forms of bias. Many of these cases involved Naming/Documentation biases, exclusionary assumptions, or semantically loaded variable and function names that were not captured by earlier labelling approaches. This suggests that prior bias labelling methodologies~\cite{huang2025bias,PNASBias2025} for code may not fully account for more granular or context-dependent bias categories. As a result, existing datasets may underestimate the prevalence of subtle biases in AI-generated code, particularly those that are embedded in developer semantics, naming conventions, or representational choices rather than explicit discriminatory logic. Our findings therefore highlight the importance of fine-grained taxonomies and contextual reasoning when auditing bias in generated code.

Proxy bias achieved the lowest detection performance across all evaluated bias categories, with a precision of 2.6\%, recall of 20\%, and F1-score of 4.7\%. This result highlights the difficulty of identifying proxy biases, as it requires reasoning about whether a seemingly neutral attribute indirectly represents a protected characteristic. Unlike more explicit forms of bias, such as Naming/Documentation or Stereotype bias, proxy bias often depends on additional contextual or domain-specific knowledge that may not be available from the code alone. For example, an attribute such as occupation could potentially act as a proxy for socioeconomic status in some contexts, but may not represent a meaningful proxy in others. As a result, LLMs may incorrectly identify proxy relationships when plausible correlations exist, leading to false positives, or fail to detect proxies when the relationship is not explicitly represented in the code. Furthermore, the limited number of proxy bias examples in the manually labelled subset (N=5) limits the reliability of category-specific performance estimates and indicates the need for more diverse benchmark datasets containing a broader range of indirect bias examples.

Overall, these findings suggest that while LLMs are effective at identifying more explicit and linguistically represented biases, detecting subtle forms of bias that require contextual reasoning remains a challenging task.

\begin{tcolorbox}[colback=gray!5, colframe=black!60, title=Key Insight: Dataset-Level Implications]
Post-hoc bias auditing extends beyond bias detection to dataset quality discovery. Even manually curated ``unbiased''` labels in existing datasets contain subtle naming and documentation biases, suggesting that prior annotation approaches may underestimate the prevalence of implicit bias in code. This indicates that ICL-based auditing can also serve as a mechanism for identifying hidden limitations in benchmark datasets.
\end{tcolorbox}

More broadly, our work contributes a shift in perspective within the literature. Existing research on bias in AI-generated code has centered on prevalence studies and mitigation during generation~\cite{ Feng2023SocialBiasCode,huang2025bias}. In contrast, we frame bias detection as a post-hoc reasoning and justification task. This distinction is increasingly relevant as AI-generated code becomes integrated into production software pipelines, where organizations require practical mechanisms to audit generated artifacts after deployment rather than only controlling generation itself.

\subsection{Increasing ICL Complexity Improves Precision More Than Recall}

\begin{tcolorbox}[colback=gray!5, colframe=black!60, title=Key Insight: Nature of Bias in AI-Generated Code]
Bias in AI-generated code is predominantly semantic rather than structural. Naming, stereotype, and demographic biases occur far more frequently than proxy or structural biases, suggesting that current CGTs primarily encode bias through developer-facing semantics such as identifiers and documentation rather than complex program logic.
\end{tcolorbox}

One notable trend is that more sophisticated prompting strategies primarily improved precision rather than overall recall. For example, Gemini ICL5 achieved the highest precision (85.69\%) and the highest code recognition accuracy (90.97\%), while recall remained relatively stable across all prompting configurations. This suggests that additional contextual examples and structured prompting as seen in prior work \cite{binkhonain2025prompts} help the model better distinguish between genuinely biased logic and benign implementation details. 

The diminishing gains across later ICL variants also suggest there may be a performance ceiling for prompt-only approaches. While richer contextual prompting improved precision, improvements between ICL3, ICL4, and ICL5 were comparatively modest. This may indicate that future improvements in bias analysis will require mechanisms beyond prompting alone, such as retrieval augmentation, external reasoning modules, symbolic analysis, or hybrid human-AI review systems.

Our findings also suggest that prompt engineering can influence how models interpret the scope of biased code. Across some cases, the identified biased code lines varied slightly between prompting strategies, typically by selecting either a smaller or larger portion of the code snippet while still capturing the same underlying mechanism of harm. This indicates that models may differ in how granularly they localize bias, even when agreement exists regarding the presence and category of bias itself. Such variation highlights the importance of evaluating not only whether a model identifies bias correctly, but also how consistently it recognizes the exact relevant code regions.

\subsection{Explanation Alignment Suggests Interpretative Reliability}

\begin{tcolorbox}[colback=gray!5, colframe=black!60, title=Key Insight: Evaluation of Explanation Quality]
Explanation quality should be evaluated as structured reasoning rather than simple correctness or similarity. High detection performance and justification alignment do not necessarily guarantee consistent reasoning structure or localization granularity, highlighting the need for more nuanced evaluation frameworks for CGT explanations.
\end{tcolorbox}

Across Gemini configurations, justification similarity scores remained consistently near 80\%, indicating moderate-to-high alignment between LLM-generated explanations and human-authored justifications. Suggesting that the model was often capable not only of identifying biased code but also articulating why the code was considered biased.

This finding is important because explainability remains a central challenge in responsible AI workflows. From a human factors perspective, explanation quality directly influences whether developers trust and appropriately act upon AI-generated recommendations. High detection accuracy alone is insufficient if developers cannot understand why a code fragment has been flagged or determine whether the explanation is applicable within their development context. Consequently, explanation quality becomes a critical component of responsible GenAI adoption rather than merely an auxiliary evaluation metric. Prior work in NLP bias detection has emphasized interpretable explanations and token-level rationales~\cite{DBias2022, BiasAwareAgent2025}, but these approaches primarily focus on natural language. Our findings demonstrate that similar explanation capabilities may extend to executable code, where bias is often embedded in logic, conditions, and program structure rather than surface language.

However, the gap between recognition accuracy and justification similarity also reveals an important limitation. While the model frequently identified the correct biased region, its explanation did not always fully align with expert reasoning. In cases where justification similarity differed by more than 20\%, manual evaluation revealed that the generated explanations often remained semantically consistent with the intended bias category, but used more descriptive or generalized language than expert-authored explanations. Importantly, these explanations were typically not outside the conceptual scope of the identified bias. Hong et al. shared a similar experience in their work with LLMs from a linguistic perspective where they highlighted LLM outputs labelled as incorrect commonly consisted of acceptable alternative responses rather than strict errors \cite{hong2025litex}. This indicates that disagreement scores may partially reflect differences in explanatory framing and verbosity rather than incorrect reasoning which can be expected for stochastic models but can be unfairly be penalized \cite{jiang2022investigating}. 

Similarly, variations in code recognition often stemmed from differences in how much of the surrounding code context the model included when identifying biased logic. In many cases, the model selected slightly smaller portions of the snippet while still correctly identifying the central biased mechanism. This finding highlights an important nuance in evaluating explanation reliability: differences in localization granularity or explanatory detail do not necessarily imply conceptual disagreement regarding the underlying bias.

These findings have broader implications for explainable AI research in software engineering. Existing fairness evaluations often treat explanation alignment as a binary correctness task but recent pushes in AI explainability have argued it should be evaluated as a multi-faceted concept \cite{nauta2023anecdotal}. Likewise, our results support this in that bias justification similarity in code analysis may involve multiple dimensions, as also seen in the medical domain \cite{hu2024towards} including conceptual agreement, contextual scope, granularity, and descriptive specificity. As a result, future evaluation methodologies may need to distinguish between substantive reasoning disagreement and variations in explanation style.

At the same time, our findings reinforce concerns from explainable AI literature regarding post-hoc rationalization. While LLM-generated explanations show promise as developer support tools, they should likely be treated as assistive rather than authoritative explanations. Human oversight remains necessary to validate nuanced fairness judgments and contextual implications.

Importantly, our findings do not suggest that developers should blindly trust LLM-generated fairness assessments. While the consistently high recall indicates that the model is effective at identifying potentially problematic code, the lower precision demonstrates that some flagged cases require human verification. Consequently, post-hoc bias auditing is best viewed as a human-in-the-loop decision-support mechanism that prioritizes surfacing potential risks while leaving the final fairness judgment to developers.

\subsection{Implications for Responsible AI and Software Engineering}

Our findings offer various implications for responsible AI practices in software engineering.

\ding{42} \textbf{Bias Auditing:} Bias auditing may become an increasingly necessary stage of the software development lifecycle as CGT adoption grows, particularly in high-stakes domains such as healthcare where concerns around biased or unsafe outputs are becoming more apparent \cite{alu2026auditing}. Existing software engineering workflows already incorporate static analysis, security scanning, and testing pipelines. Our results suggest that bias auditing could emerge as an analogous layer for identifying socially or structurally harmful logic before deployment.

\ding{42} \textbf{Responsible Human Adoption of GenAI:} Rather than replacing existing review practices, post-hoc bias auditing should complement human code review by providing developers with transparent explanations of potentially harmful logic. This supports responsible adoption of GenAI by encouraging developers to critically evaluate AI-generated code instead of assuming functional correctness implies ethical acceptability.

\ding{42} \textbf{Supporting Trust Calibration:} Effective adoption of GenAI requires developers to appropriately calibrate their trust in AI-generated recommendations. Our results suggest that high-recall bias auditing may help developers identify potentially problematic code, while explanation alignment provides the contextual information needed to evaluate whether flagged issues warrant further investigation. This encourages informed reliance rather than blind acceptance or outright dismissal of AI-generated feedback.

\ding{42} \textbf{CGTs as Socio-Technical Systems:} Our findings reinforce the need to treat code generation systems as socio-technical systems rather than purely technical productivity tools. Bias in generated code is not limited to overt discrimination; it can also reinforce dominant platforms, encode exclusionary assumptions, or reproduce historical inequities embedded in training data. This aligns with prior work arguing that purely technical frameworks often fail to account for the socio-technical dynamics surrounding AI systems and historical inequities \cite{selbest2019}. Consequently, evaluating AI-generated code solely on correctness or efficiency may overlook important fairness concerns.

\ding{42} \textbf{Reducing Barriers to Fairness:} The effectiveness of post-hoc ICL analysis suggests that lightweight auditing approaches may be viable even for organizations without access to proprietary model internals or expensive retraining pipelines. This lowers the barrier for incorporating fairness-oriented review practices into development workflows and increases the practical applicability of our approach. This is particularly important given prior work highlighting the practical and translational barriers associated with deploying fairness interventions in real-world AI systems \cite{liu2023towards}.

\ding{42} \textbf{Moving Beyond Code Correctness:} Existing CGT research has primarily emphasized productivity, acceptance, and correctness. In contrast, our study introduces explanation reliability and interpretative alignment as important dimensions for evaluating AI-generated code. This expands current evaluation techniques by suggesting that future CGT benchmarks should assess not only whether generated code functions correctly, but also whether the reasoning surrounding potentially harmful logic can be transparently explained and audited.

More broadly, our findings suggest that the future of responsible AI-assisted software engineering is unlikely to rely solely on improving code generation models themselves. More broadly, our findings suggest that responsible adoption of GenAI in software engineering may depend as much on effective human-centered auditing workflows as on improvements in code generation itself. Rather than viewing explainability and bias auditing as optional post-processing steps, they may become core mechanisms for helping developers understand, question, and validate AI-generated code before deployment. 
In this view, explainability, calibrated trust, and human oversight become essential components of responsible GenAI adoption alongside functional correctness and software quality.

\subsection{Recommendations}
Based on the empirical insights and trade-offs uncovered across our research questions, we formulate targeted, actionable recommendations for software engineering stakeholders.

\subsubsection{Recommendations for Practitioners and Industry:} Organizations using CGTs should integrate post-hoc bias auditing into existing software review pipelines alongside security and static analysis tools. Our findings show that AI-generated code may encode not only demographic bias, but also proxy, structural, and ecosystem-level biases that are difficult to detect through traditional testing alone. This can apply to governance frameworks as well, since policy and regulatory evaluation processes often rely on standardized compliance checks that may overlook how AI systems propagate systemic inequities through indirect variables, institutional assumptions, or downstream deployment contexts.

Practitioners should treat LLM-generated bias explanations as auxiliary evaluation material rather than authoritative ground truth.  Although our framework proved that alignment scores remain consistently high ($\approx$ 90\%), differences in contextual scope and explanatory framing remained common, reinforcing the need for human oversight in high-impact domains. Human-in-the-loop review should be mandatory when deploying automated audits in safety-critical, legally compliant, or socially sensitive software domains.


\subsubsection{Directions for Software Engineering Researchers:} Future work should move beyond explicit demographic bias and continue investigating underexplored forms of bias in AI-generated code. Researchers should also consider that additional bias patterns or harmful interactions may exist outside the contexts captured in our dataset, including biases that emerge only under different application domains, user populations, system integrations, or real-world deployment conditions.

Researchers should also develop more nuanced evaluation methodologies for explanation alignment. Our findings suggest that disagreement often reflects differences in granularity or framing rather than fundamentally incorrect reasoning. Future benchmarks must employ semantic entailment metrics rather than surface-level lexical comparisons to prevent valid reasoning alternatives from being incorrectly classified as logical failures.

Given the clear \emph{diminishing returns of prompt-only scaling}, researchers must transition away from isolated text-instruction engineering. Future work should pioneer hybrid architectures that explicitly couple LLM reasoning modules with Abstract Syntax Tree (AST) parsing, symbolic execution, and Retrieval-Augmented Generation (RAG) to successfully break through current open-weights performance ceilings.

Researchers should work on constructing standardized, multi-label benchmarks, taxonomies, and human-verified datasets explicitly tailored for source code files. While fairness research remains deeply mature within natural language processing (NLP), the software engineering domain lacks the foundational, open-access evaluation infrastructure required to automated code-level bias detection.

\subsubsection{Recommendations for Academia and Education:} Software engineering curricula should increasingly incorporate Fairness, Accountability, Transparency, and Ethics (FATE) into AI-assisted programming education. Students should be trained  not only to execute prompt engineering for generation, but also to critically audit synthesized code blocks for subtle, linguistically embedded demographic assumptions and subtle bias propagation.


\section{Threats to Validity}
We evaluate the threats to validity across four distinct dimensions: 

\textbf{Construct Validity}: Bias is deeply context-dependent, culturally situated, and socially constructed; therefore, expert annotations and human-authored justifications involve an element of subjective interpretation. We mitigated this drift by enforcing strict structural templates (Trigger, Mechanism, Impact) and executing rigorous, multi-stage inter-rater calibration loops. Standardized Impact templates improve reproducibility but may introduce lexical overlap between expert and model explanations, potentially increasing JSS values. Therefore, JSS should be interpreted as alignment with expert-structured reasoning components rather than a standalone measure of explanation quality. Although our taxonomy, prompt refinement process, and justification framework were informed by prior literature, their design involved methodological decisions by the authors that may introduce subjectivity into the evaluation.

\textbf{Internal Validity}: A critical threat in empirical research when using LLMs is data contamination. Because we build upon an existing evaluation corpus~\cite{huang2025bias}, closed-source LLMs (e.g., Gemini) or newly updated open-weights base layers may have seen the baseline data during continuous pre-training. High baseline performance might partially reflect token memorization. 

\textbf{Conclusion Validity}: To ensure reproducibility, we reported metrics averaged over multiple independent executions ($N=3$). Also, the current evaluation primarily examined textual and logical reasoning rather than downstream behavioral impacts during execution. A model may correctly identify potentially biased logic without accurately predicting the real-world harms produced once the software is deployed and interacts with live databases, external APIs, or human end-users.

\textbf{External Validity}: First, our model selection, while capturing a diverse cross-section of proprietary and open-source models, it might be possible that smaller models or alternative architectures may exhibit different reasoning behaviors. Although using a fixed ICL configuration enables consistent comparison across LLMs, the optimal ICL strategy may vary across model architectures. Therefore, the selected configuration may not represent the best-performing setting for every evaluated model. Future work should investigate the full ICL configuration space independently for each LLM. Second, our experimental pipeline is restricted to isolated Python code functions. The discovered taxonomy and prompt behaviors may not fully generalize to statically-typed languages (e.g., Java, C++), larger software repositories, or more complex real-world systems where biases are likely to emerge through complex dependencies. Similarly, while our taxonomy captures a broader range of bias categories than prior work, additional undocumented bias types may still exist outside the scope of our current dataset. 


\section{Conclusion}

This work explored bias in AI-generated code through three research questions focused on the types of biases that emerge in generated code, the effectiveness of ICL for post-hoc bias detection, and the extent to which LLM-generated explanations align with expert-authored justifications. Our findings show that AI-generated code contains a broader range of biases than what has primarily been explored in prior work, including proxy, representational, and documentation/naming biases in addition to demographic and stereotype bias.

We also demonstrate that post-hoc bias detection using ICL is feasible without requiring model fine-tuning or access to proprietary model internals, with consistently strong bias identification and code recognition performance across prompting strategies. Further, our results suggest that LLM-generated explanations can achieve moderate-to-high interpretative alignment with expert-authored justifications, while still exhibiting variation in contextual scope, explanatory detail, and granularity.

Overall, this work contributes a broader characterization of bias in AI-generated code, an empirical evaluation of LLM-based post-hoc bias auditing, and an analysis of explanation reliability for bias detection in source code. These findings help move the discussion beyond whether bias exists in AI-generated code toward understanding how bias manifests, how it can be identified, and how it can be more transparently explained within AI-assisted software engineering workflows.

\section{Data Availability}

All related data, scripts, and the replication package have been made available \cite{inclusiveai}.








\bibliographystyle{ACM-Reference-Format} 
\bibliography{references}






\end{document}